\documentclass[floatfix,a4paper,tightenlines,amsmath,amssymb,nofootinbib,showpacs,notitlepage,showkeys,prd,twocolumn,10pt]{revtex4-1}

\usepackage[utf8]{inputenc}
\usepackage{graphicx}
\usepackage{hyperref}
\usepackage{slashed}

\newcommand{\de}{\delta}
\newcommand{\eref}[1]{Eq.~(\ref{#1})}
\newcommand{\tref}[1]{Tab.~\ref{#1}}
\newcommand{\fref}[1]{Fig.~\ref{#1}}
\newcommand{\nnnl}{\nonumber\\}	

\DeclareMathOperator{\Tr}{Tr}
\DeclareMathOperator{\sumint}{\mathchoice
  {\ooalign{\raisebox{.08\height}{\scalebox{.85}{$\displaystyle\sum$}}\cr\hidewidth$\displaystyle\int$\hidewidth\cr}}
  {\ooalign{\raisebox{.24\height}{\scalebox{.6}{$\textstyle\sum$}}\cr\hidewidth$\textstyle\int$\hidewidth\cr}}
  {\ooalign{\raisebox{.24\height}{\scalebox{.6}{$\scriptstyle\sum$}}\cr$\scriptstyle\int$\cr}}
  {\ooalign{\raisebox{.24\height}{\scalebox{.6}{$\scriptstyle\sum$}}\cr$\scriptstyle\int$\cr}}
}

\allowdisplaybreaks 

\begin{document}

\title{Dense two-color QCD from Dyson-Schwinger equations}

\author{Romain Contant}
\email{contant.romain.phy@gmail.com}
\affiliation{Institute of Physics, University of Graz, NAWI Graz, Universit\"atsplatz 5, 8010 Graz, Austria}

\author{Markus Q.~Huber}
\email{markus.huber@physik.jlug.de}
\affiliation{Institut f\"ur Theoretische Physik, Justus-Liebig--Universit\"at Giessen, Heinrich-Buff-Ring 16, 35392 Giessen, Germany}

\date{\today}

\begin{abstract}
We investigate quantum chromodynamics with two colors at nonvanishing density using Dyson-Schwinger equations.
Lattice methods do not have a complex action problem in this theory.
Thus, we can benchmark our results and the effect of truncations directly by comparing with the corresponding lattice results.
We do so for the gluon propagator, the chiral condensate, and the quark number density and test variations of the employed truncation to improve the agreement.
Finally, we compare the effect of a truncation on the chiral and confinement/deconfinement transitions in the phase diagrams of QCD and QCD with the gauge groups $SU(2)$ and $G_2$.
\end{abstract}



\maketitle

\section{Introduction}

Strongly interacting matter possesses a rich phase structure.
At vanishing chemical potential, there is a crossover from a hadron dominated phase at low temperatures to the so-called quark-gluon plasma phase at high temperatures.
This is well established, for example, by lattice simulations \cite{Borsanyi:2010bp,Bazavov:2011nk,Bhattacharya:2014ara,Bazavov:2014pvz}.
Switching on a chemical potential, though, the picture becomes less clear.
The otherwise successful method of Monte Carlo lattice simulations faces a complex action problem that currently restricts calculations to a baryon chemical potential of $\mu_B\lesssim 2T$~\cite{deForcrand:2010ys}.

In that regime, no critical endpoint has been found by lattice simulations and it is up to other methods to explore the regime beyond that.
For cold and dense strongly interacting matter, a rich phase structure is expected, but the details are still not clear besides that at high enough densities a color superconducting phase will appear \cite{Alford:1997zt,Alford:1998mk,Berges:1998rc,Alford:1999pa,Buballa:2003qv,Rischke:2003mt,Alford:2007xm}.

Thus, to explore high densities, alternative approaches to lattice simulations are required.
One such method is functional equations; for general reviews, see, e.g., \cite{Berges:2000ew,Roberts:2000aa,Alkofer:2000wg,Pawlowski:2005xe,Fischer:2006ub,Gies:2006wv,Schaefer:2006sr,Binosi:2009qm,  Braun:2011pp,Maas:2011se,Eichmann:2016yit,Sanchis-Alepuz:2017jjd,Huber:2018ned}, and for their application to the phase diagram of quantum chromodynamics (QCD), e.g., \cite{Braun:2007bx,Braun:2010cy,Pawlowski:2010ht,Fischer:2011pk,Fischer:2011mz,Fischer:2012vc,Fischer:2013eca,Qin:2013ufa,Haas:2013hpa,Xin:2014ela,Fischer:2014ata,Gao:2015kea,Drews:2016wpi,Fukushima:2017csk,Fischer:2018sdj,Isserstedt:2019pgx,Gunkel:2019xnh,Fu:2019hdw,Hajizadeh:2019qrj}.

They provide access to the full phase diagram but two things need to be dealt with. First, from a technical point of view, nonvanishing density and temperature complicate the equations due to the loss of manifest Lorentz invariance. Second, being exact equations, they can only be solved once approximations are made because they form an infinitely large system of equations.
In the vacuum, several studies indicate that the system has favorable convergence properties \cite{Cyrol:2016tym,Huber:2016tvc,Cyrol:2017ewj,Huber:2017txg,Corell:2018yil,Huber:2018ned}.
However, in the medium, the corresponding level of truncation is not reached yet, and once it is, it is not guaranteed that the convergence properties remain the same.
Thus, for the time being, calculations in the medium operate at a more basic level involving phenomenological models for interactions.

At vanishing chemical potential, 'intermediate' (gauge dependent) results, like propagators and vertices, as well as 'final' (gauge invariant) results, like condensates, of functional equations can be compared against results from lattice calculations.
Thus, truncations can be assessed directly.
At nonvanishing density, the possibilities for such comparisons are very limited.
One way is to switch to a theory which does not suffer from a complex action problem.
This is the path we follow here.
We will solve Dyson-Schwinger equations (DSEs) of QC$_2$D \cite{Kogut:2000ek}, which is QCD with the gauge group $SU(2)$ instead of $SU(3)$.
For an even number of quark families it has no complex action problem.
Lattice calculations were performed for this theory \cite{Kogut:2002cm,Hands:2006ve,Hands:2010gd,Hands:2011ye,Hands:2012yy,Cotter:2012mb,Boz:2013rca,Braguta:2016cpw,Boz:2018crd,Wilhelm:2019fvp}, and we will compare results for the gluon propagator, the quark number density, and the chiral condensate.
QC$_2$D was also investigated with various continuum methods, e.g., \cite{Sun:2007fc,Brauner:2009gu,He:2010nb,Andersen:2010vu,Strodthoff:2011tz,vonSmekal:2012vx,Strodthoff:2013cua,Buscher:2014ixt,Khan:2015puu,Contant:2017gtz,Suenaga:2019jjv}.
Furthermore, it has attracted attention \cite{Vujinovic:2014ioa,Arthur:2016ozw,Arthur:2016dir,Drach:2017btk,Lee:2017uvl,Drach:2017jsh,Vujinovic:2018nko} as a potential theory for a composite Higgs in the context of technicolor \cite{Cacciapaglia:2014uja}.

A second alternative to QCD is QCD with the gauge group $G_2$ for which lattice simulations do not have a sign problem either \cite{Holland:2003jy,Pepe:2006er}.
We will also consider this theory shortly in Sec.~\ref{sec:results} where we compare the phase diagrams of QCD and the two QCD-like theories with respect to chiral symmetry and confinement.

The spectrum of QC$_2$D differs from real QCD as it does not have fermionic baryons.
Instead, it contains color-neutral diquarks which can condense and lead to the existence of a transition from a Bose-Einstein condensate (BEC) to a Bardeen-Cooper-Schrieffer (BCS) phase.
Diquarks can be taken into account with the Nambu-Gor'kov formalism \cite{Rischke:2003mt}.
We will not include diquarks at this point as this would increase the complexity of the calculations.
We will, however, come back to them in Sec.~\ref{sec:summary} and shortly touch upon their impact on our calculations.
Calculations with diquarks and a similar setup can be found in Ref.~\cite{Buscher:2014ixt}.

In the next section we detail our setup.
The results are presented in Sec.~\ref{sec:results}, and we summarize and provide an outlook in Sec.~\ref{sec:summary}.

\section{Setup}
\label{sec:setup}

\begin{figure}[tb]
 \begin{center}
 \includegraphics[width=0.45\textwidth]{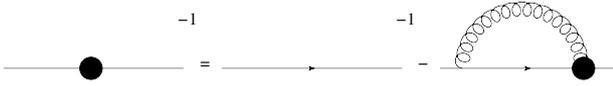}
 \caption{Quark propagator DSE. Quantities with a blob are fully dressed, as are internal propagators. Continuous/wiggly lines denote quarks/gluons.}
 \label{fig:DSEqqb}
 \end{center}
\end{figure}

\begin{figure}[tb]
 \begin{center}
 \includegraphics[width=0.45\textwidth]{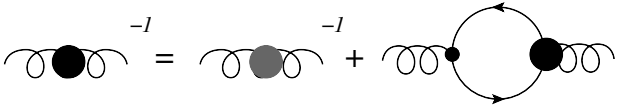}
 \caption{The gluon propagator DSE is split into a quenched part (gray blob) and the quark loop. The former is determined from quenched lattice results.}
 \label{fig:AADSEApprox}
 \end{center}
\end{figure}

We solve the gluon and quark propagator DSEs using a model for the quark-gluon vertex and lattice input for the quenched gluon propagators.
The equations are shown in Figs.~\ref{fig:DSEqqb} and \ref{fig:AADSEApprox}.
As is made explicit in \fref{fig:AADSEApprox}, the quark loop is split off in the gluon propagator DSE.
We approximate the rest by quenched lattice results and calculate the quark loop explicitly.
This approximation, developed in a series of works \cite{Nickel:2006vf,Fischer:2009wc,Fischer:2012vc}, neglects only indirect quark contributions, but the obvious advantage is that neither gluonic vertices are required nor two-loop contributions need to be calculated,
both of which are quantitatively relevant \cite{Huber:2017txg,Huber:2018ned}.
This particular truncation was already used for calculations of the $N_f=2$ \cite{Fischer:2011pk,Fischer:2011mz,Fischer:2012vc}, $N_f=2+1$ \cite{Fischer:2011mz,Fischer:2014ata}, and $N_f=2+1+1$ \cite{Fischer:2014ata} QCD phase diagrams, of baryon number fluctuations \cite{Isserstedt:2019pgx} and of mesons at nonvanishing chemical potential \cite{Gunkel:2019xnh}.

The quark propagator DSE reads
\begin{widetext}
\begin{align}\label{eq:quarkDSE}
S^{-1}(p)&=Z_2 S_0^{-1}(p) +Z_{1F} C_{F} g^{2}\sumint_q \gamma_{\mu} S(q) \Gamma_{\nu}(p-q;-p,q) D_{\mu \nu}(p-q)
\end{align}
\end{widetext}
with the momenta $p=(\vec{p},p_4)$ and $q=(\vec{q},q_4)$.
The quark and gluon propagators have Matsubara frequencies $p_4=(2n_p+1)\pi T$ and $p_4=2n_p\pi T$, respectively, with $n_p\in \mathbb{Z}$.
The integration and Matsubara summation are abbreviated as $\sumint_q \equiv T \sum_{n_q \in \, \mathbb{Z}} \int d^3 \vec{q} \, / \, (2\pi)^3 $.
$C_F=(N_c^2-1)/(2N_c)$ is the quadratic Casimir of the fundamental representation.
$S(p)$ is the full quark propagator parametrized as
\begin{align}\label{eq:quarkprop}
S^{-1}(p) &= i \vec{p} \vec{\gamma} A(p) + i(p_4+i\,\mu) \gamma_{4} C(p)\nnnl
&+ B(p) +  i p_4 \gamma_{4} \vec{p} \vec{\gamma} D(p).
\end{align}
In our calculations, we set $D(p)=0$.
At vanishing density, we tested its quantitative influence and found at most a change of the order of $0.0001$ on the other dressing functions \cite{Contant:2017gtz}.
$S_0^{-1}(p)=i \vec{p} \vec{\gamma} + i (p_4+i\,\mu) \gamma_{4} + Z_m m$ is the inverse bare quark propagator, where $Z_m$ and $Z_2$ are the quark mass and wave function renormalization constants, respectively.
$m$ is the renormalized current quark mass at the renormalization point.
Finally, $Z_{1F}$ is the quark-gluon vertex renormalization constant for which we use $Z_{1F}=Z_2 \widetilde{Z}_1/\widetilde{Z}_3$.
$\widetilde{Z}_1$ is the renormalization constant of the ghost-gluon vertex which is finite in Landau gauge \cite{Taylor:1971ff} and we choose it as $1$.

The quark-gluon vertex is approximated by the following model \cite{Fischer:2009wc}:
\begin{align}
\label{eq:qug}
&\Gamma_{\nu}(q;p,l) =\widetilde{Z}_3 \gamma_{\mu} \Gamma_\text{mod}(x) \nnnl
&\quad\times\Bigg(\frac{A(p^2) + A(l^2)}{2} \delta_{\mu, i}+\frac{C(p^2) + C(l^2)}{2} \delta_{\mu, 4} \Bigg),\\
&\Gamma_\text{mod}(x) = \frac{d_{1}}{\left(x+d_2\right)} \nnnl
&\quad+ \frac{x}{\Lambda^2 + x} \left(\frac{\alpha(\mu)\beta_{0}}{4 \pi}\textrm{ln}\left(\frac{x}{\Lambda^2} + 1\right)\right)^{2 \delta}.
 \label{eq:qglvertModel}
\end{align}
$p$ and $l$ are the antiquark and quark momenta, respectively, and $q$ is the gluon momentum.
$x$ depends on the equation in which the vertex model is used.
This is necessary to maintain multiplicative renormalizability with this model \cite{Fischer:2003rp}.
$x$ is $(p^2+l^2)$ in the gluon propagator DSE and $q^2$ in the quark propagator DSE.
The function $\Gamma_\text{mod}(x)$ contains the logarithmic running in the perturbative regime.
Its strength in the IR is determined by the value of the parameter $d_1$.
$\Lambda$ and $\alpha(\mu)$ are fit parameters taken from the gluon propagator fit; see below.
The parameter $d_2$ is fixed at $0.5\,\text{GeV}^2$.
$\delta$ is the one-loop anomalous dimension of the ghost propagator given by $\de=-9N_c/(44N_c-8N_f)$ and $\beta_0=(11N_c - 2N_f)/3$ is the lowest coefficient of the QCD beta function.
The model has twice the anomalous dimension of the quark-gluon vertex which is necessary to obtain the correct anomalous dimensions of the propagators \cite{Fischer:2003rp} without including higher loop contributions \cite{vonSmekal:1997vx,Huber:2018ned}.
An implicit temperature and chemical potential dependence enters via the quark propagator dressing functions $A(p^2)$ and $C(p^2)$.

The gluon propagator DSE reads
\begin{widetext}
\begin{align}
 D_{\mu\nu}^{-1}(p)=\left[D_{\mu\nu}^{\text{YM}}(p)\right]^{-1}-\frac{g^{2}}{2} N_f Z_{\text{1F}}\sumint_q \, \Tr\left[\gamma_{\mu} \, S(p+q) \,\Gamma_{\nu}(-p;-q,p+q) \, S(q)\right].
\end{align}
\end{widetext}
$D_{\mu\nu}^{\text{YM}}$ contains all gluonic diagrams.
In the medium, the propagator splits into parts transverse and longitudinal with respect to the heat bath,
\begin{align}
 D_{\mu\nu}(p)=&D_{L,\mu\nu}(p)+D_{T,\mu\nu}(p)\nnnl
 =&P^L_{\mu\nu}(p)\frac{Z_L(p^2)}{p^2} + P^T_{\mu\nu}(p)\frac{Z_T(p^2)}{p^2},
\end{align}
with
\begin{align}
 P^T_{\mu\nu}(p)&=(1-\delta_{\mu4})(1-\delta_{\nu4})\left(\delta_{\mu\nu}-\frac{p_\mu p_\nu}{\vec{p}^2}\right),\\
 P^L_{\mu\nu}(p)&=P_{\mu\nu}-P^T_{\mu\nu}(p),\\
 P_{\mu\nu}&=\delta_{\mu\nu}-\frac{p_\mu p_\nu}{p^2}.
\end{align}
For the gluonic part, the dressing functions $Z^\text{YM}_T$ and $Z^\text{YM}_L$ are fitted by \cite{Fischer:2010fx}
\begin{align}\label{eq:ZTL}
Z^\text{YM}_{ T/L }(p^2)=\frac { x }{ (x+1)^2 } \Bigg( \left( \frac { c/\Lambda^2 }{ x+a_{ T/L } }  \right) ^{ b_{ T/L } }+\nnnl
 x\left( \frac { \alpha (\mu )\beta _{ 0 } }{ 4\pi  } \textrm{ln}(x+1) \right) ^{ \gamma  } \Bigg)
\end{align}
with $x=p^2/\Lambda^2$.
We only fit the lowest Matsubara frequency and use for higher Matsubara frequencies $Z^\text{YM}_{T/L}(\vec{p}^2, p_4^2)=Z^\text{YM}_{T/L}(\vec{p}^2+p_4^2,0)$, which is a good approximation according to lattice results \cite{Fischer:2010fx}.
The parameters are $c = 11.5\,\text{GeV}^2$ and $\Lambda = 1.4\,\text{GeV}$.
$\alpha(\mu)=g^2/4\pi=0.3$ is used throughout all calculations.
The fits also set the scale in our calculations.
$\gamma=(-13N_c+4N_f)/(22N_c-4N_f)$ is the one-loop anomalous dimension of the gluon propagator.

Temperature dependence enters via the fit parameters $a_{T/L}$ and $b_{T/L}$.
They are fitted to the lattice results of Refs.~\cite{Fischer:2010fx,Maas:2011ez}.
To have a smooth behavior and access to all temperatures, the parameters are fitted in terms of $t = \frac{T}{T_c}$ \cite{Contant:2017gtz},
\begin{align}
a_T &= \left \{
	\begin{array}{l@{}@{\hspace{1em}}lr}
		0.46 \ge t  :& 1.41 + 0.43 t \\
		1 \ge t \ge 0.46 :& 1.52 + 0.20 t \\
		t \ge 1 :& 3.60 - 1.88t
	\end{array} \right. \\
b_T&=\left\{
	\begin{array}{l@{}@{\hspace{1em}}lr}
		0.49 \ge t  :& 2.20 + 0.07 t \\
		1 \ge t \ge 0.49 :& 2.43 - 0.40 t \\
		t \ge 1 :& 2.32 - 0.29t
	\end{array} \right. \\
a_L&=\left\{
	\begin{array}{l@{}@{\hspace{1em}}lr}
		0.53 \ge t  :& 1.41 - 2.09 t \\
		1 \ge t \ge 0.53 :& 0.89 - 1.51 t + 0.77 t^2 \\
		t \ge 1 :& -8.16 + 8.31t
	\end{array} \right. \\
b_L&=\left\{
	\begin{array}{l@{}@{\hspace{1em}}lr}
		0.52 \ge t  :& 2.20 - 1.82t \\
		1 \ge t \ge 0.52 :& 1.22 + 0.10 t - 0.05 t ^ 2 \\
		t \ge 1 :& -1.48 + 2.75t
	\end{array} \right. 
\end{align}

\section{Results}
\label{sec:results}

In this section we present our results for various quantities.
One of the main findings is that a straightforward extension of the $\mu=0$ setup does not lead to a satisfactory comparison with lattice results at nonvanishing chemical potential.
We argue that the quark-gluon vertex is to be blamed for that and present calculations that corroborate this.

For our calculations several setups for the parameters are used which are listed in \tref{tab:parameters}.
We explain the specific choices below where they are used for the first time.
Results along the chemical potential axis were calculated at a temperature of $47\,\text{MeV}$.

\begin{table}[bt]
 \begin{tabular}{|r|r|r|}
  \hline
    &   $d_1\, [\text{GeV}^2]$ & $m\, [\text{MeV}]$\\
  \hline
  \hline
  I & 1.02 & 188\\
  \hline
  \hline
  IIa & 15 & 1.2\\
  \hline
  IIb & 7 & 35.3\\
  \hline
  IIc & 2.6 & 70.7\\
  \hline
  \hline
  IIIa & 1.93 & 150\\
  \hline
  IIIb & 1.42 & 175\\
  \hline
  IIIc & 0.92 & 200\\
  \hline
  IIId & 0.48 & 225\\
  \hline
 \end{tabular}
\caption{Different setups that are used in this work.
The renormalization point is in all calculations $80\,\text{GeV}$.
The parameters are fixed as follows:
I: Reproduction of gluon propagator at $\mu=0$.
II: Variations of quark mass and $d_1$ fixed such that $T_c=210\,\text{MeV}$ at $\mu=0$.
III: Variations of quark mass and $d_1$ fixed such that $\mu_c=650\,\text{MeV}$ at $T\approx0$.}
\label{tab:parameters}
\end{table}

\subsection{Gluon propagator}
\label{sec:gluon_prop}

\begin{figure*}[tb]
 \begin{center}
 \includegraphics[width=0.45\textwidth]{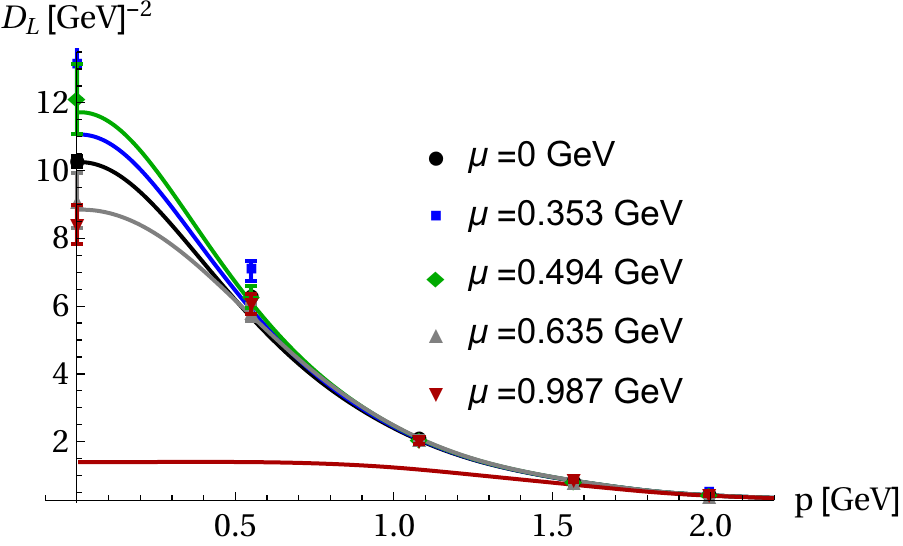}
 \includegraphics[width=0.45\textwidth]{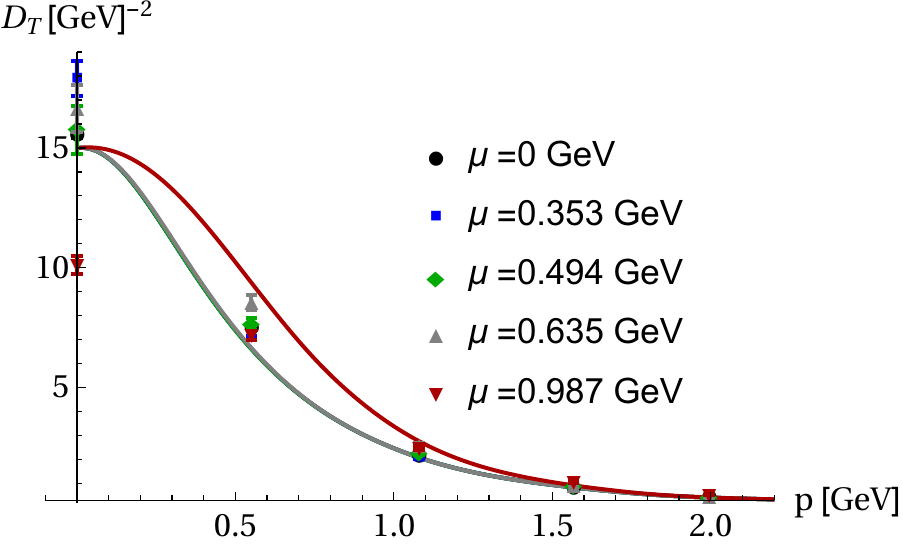}
 \caption{Left/Right: Longitudinal/Transverse gluon propagators calculated with setup I of \tref{tab:parameters}.
}
 \label{fig:glP}
 \end{center}
\end{figure*}
The truncation employed in this work takes into account gluonic effects by unquenching the gluon propagator explicitly.
The effect of finite density on the gluon propagator can thus be studied.
We start with setup I of \tref{tab:parameters} where the IR strength of the vertex $d_1$ is fixed such that we get close to the zero chemical potential gluon propagators observed in Ref.~\cite{Boz:2018crd}.
The pion mass in the lattice calculations was quite heavy with $m_\pi=717\,\text{MeV}$.
With the Gell-Mann--Oakes--Renner relation we can at least estimate the corresponding quark mass for which we use $188\,\text{MeV}$.

When solving the gluon propagator DSE, a renormalization of quadratic divergences is needed if a hard UV cutoff is employed.
Very often, a generalization of the Brown-Pennington projector \cite{Brown:1988bn} is employed.
Here, in order to compare with lattice results, for setup I we use a mass counter term $C_\text{sub}/p^2$ which is fixed by a second renormalization condition \cite{Collins:2008re,Meyers:2014iwa,Huber:2017txg,Huber:2018ned}.
Various other methods exist, but they have mostly been employed only in vacuum calculations see, e.g., \cite{Fischer:2002eq,Fischer:2002hn,Maas:2005hs,Fischer:2005en,Cucchieri:2007ta,Aguilar:2009ke,Huber:2012kd,Huber:2012zj,Huber:2014tva,Aguilar:2016vin}.

The resulting longitudinal gluon propagators are shown in \fref{fig:glP}.
As one can see in the comparison to lattice results, the agreement down to $1\,\text{GeV}$ is satisfactory.

\begin{figure}[tb]
 \begin{center}
 \includegraphics[width=0.5\textwidth]{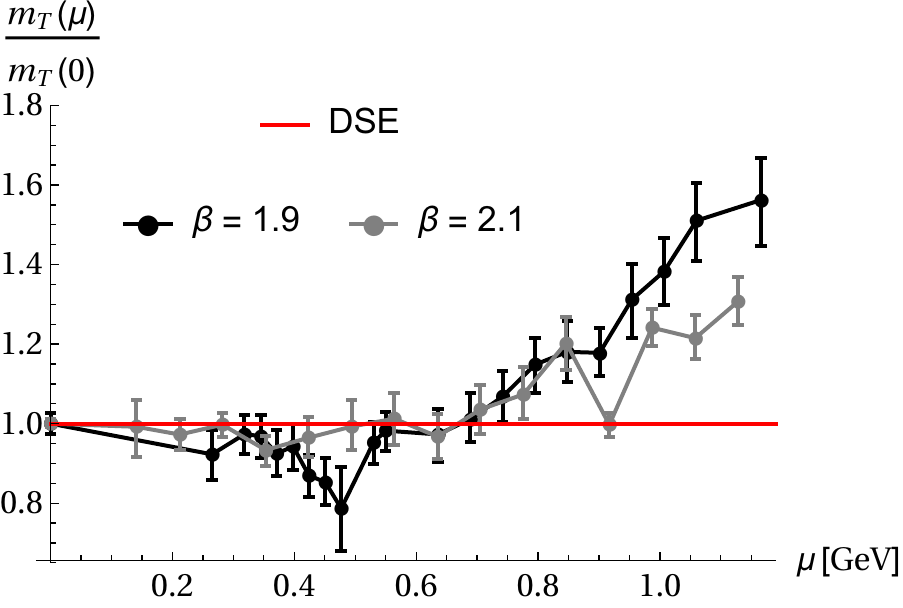}
 \caption{Transverse screening mass normalized by the vacuum value.
 In our truncation it is independent of chemical potential.
 Lattice results are from Ref.~\cite{Boz:2018crd}.
}
 \label{fig:screen_mass_T}
 \end{center}
\end{figure}

Interesting quantities to monitor the reaction of the propagators on external parameters like temperature or chemical potential are the longitudinal and transverse screening masses, $m_L$ and $m_T$, respectively.
They are defined as
\begin{align}
 m_{L/T}=\frac{1}{\sqrt{D_{L/T}(0)}},
\end{align}
where $D_{L/T}(p^2)$ is the scalar part of the corresponding propagators.
It should be noted that these screening masses are in general not the same as pole masses, which not even need to exist.
Of particular interest is the longitudinal screening mass which shows a distinct temperature dependence at zero chemical potential \cite{Fischer:2010fx,Maas:2011ez,Cucchieri:2012nx,Silva:2013maa,Boz:2018crd}.
With the truncation employed here, the transverse screening mass is completely determined by \eref{eq:ZTL} and independent of $\mu$, because the quark-loop contribution vanishes at zero momentum in this case.
The transverse screening mass is shown in \fref{fig:screen_mass_T}.
The constant transverse screening mass appears to be an acceptable approximation when compared with the lattice data \cite{Boz:2018crd} at least for chemical potential below $800\,\text{MeV}$.
However, the observed increase for larger chemical potential depends on the lattice spacing and is thus at least affected by the discretization if not a complete lattice artifact \cite{Boz:2018crd}.
Taking the indications of the small dependence of the gluon propagator on chemical potential as working assumption offers the interesting possibility of approximating the gluon propagator as independent of chemical potential.
We will explore this possibility below.

\begin{figure*}[tb]
 \begin{center}
 \includegraphics[width=0.45\textwidth]{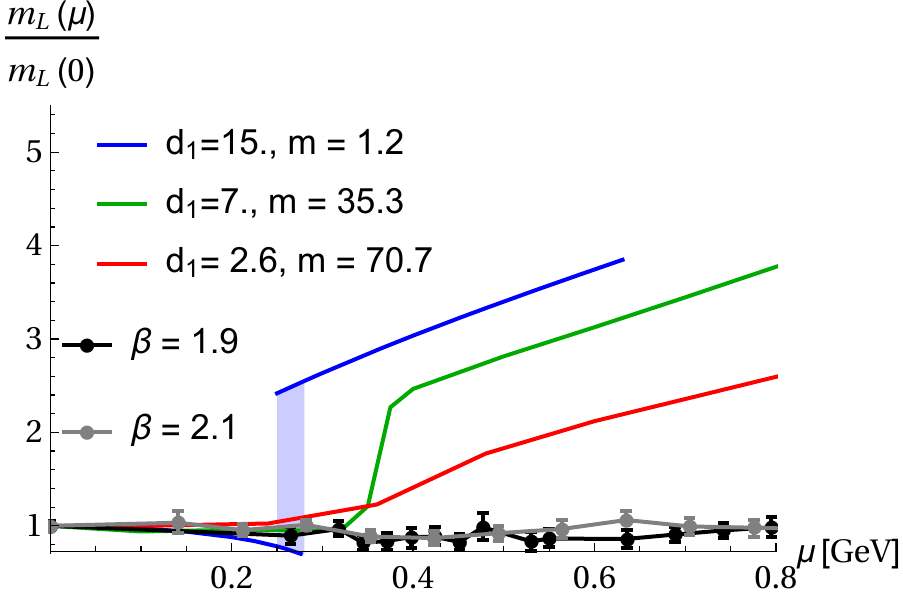}\hfill
 \includegraphics[width=0.45\textwidth]{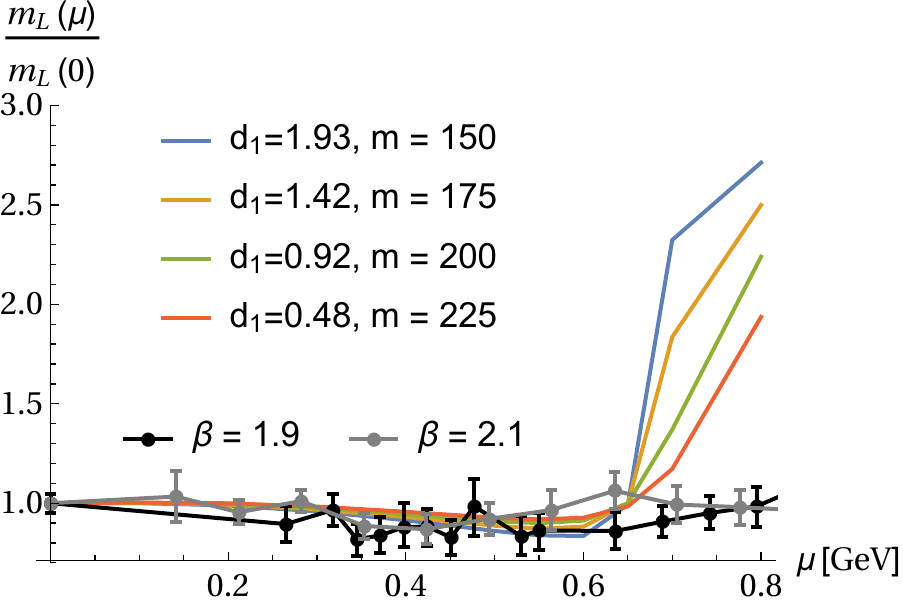}
 \caption{Left/Right: Longitudinal screening mass normalized by the vacuum value calculated with setups II/III of \tref{tab:parameters}.
}
 \label{fig:screen_mass_L}
 \end{center}
\end{figure*}

We calculate the longitudinal screening mass in two setups.
We vary the quark masses and fix $d_1$ such that the chiral crossover transition at zero chemical potential is at $210\,\text{MeV}$.
This corresponds to setup II in \tref{tab:parameters}.
The resulting longitudinal screening mass is shown in the left plot of \fref{fig:screen_mass_L}.
For the lowest quark mass, $m=1.2\,\text{MeV}$, we find a first order transition between $\mu=250\,\text{MeV}$ and $300\,\text{MeV}$.
Increasing the quark mass moves the transition to higher values of the chemical potential and smooths it to a crossover.
However, the lattice results do not show a transition in the screening mass in this region at all.
The silverblaze point, where a first order transition occurs, is located in this interval, but it is not reflected in the screening masses \cite{Boz:2018crd}.

A physically better choice for fixing $d_1$ is to use a condition from the chemical potential axis.
Lattice results indicate that the longitudinal screening mass also remains largely unaffected by chemical potential.
Only at high chemical potential a small increase is seen, which, however, might be affected by the same problems of discretization artifacts and statistics as the transverse screening mass.
Thus, with setup III of \tref{tab:parameters} we want to see if we can push the transition to higher chemical potential by varying the quark masses again and fixing $d_1$ such that the increase starts at $\mu=650\,\text{MeV}$, which is a conservative estimate for the beginning of this increase.
It should be noted that the onset of an increase is clearer for the transverse screening mass.
The quark masses are now considerably higher and consequently the setup would lead to a totally different crossover line between the hadronic regime and the quark-gluon plasma.
However, there is anyway no reason to expect that the vertex is independent of temperature and chemical potential, so having a different $d_1$ at low temperatures is to be expected.
Also, for real QCD, it was seen that fixing $d_1$ via the pion in the vacuum or via the transition temperature at $\mu=0$ yields different values \cite{Fischer:2014ata}.

The results for the longitudinal screening mass for setup III are shown in the right plot of \fref{fig:screen_mass_L}.
Increasing the quark mass lowers the steepness of the increase after the transition.
In addition, the screening mass below the transition flattens.

As mentioned above, it is not settled if the rise of the screening masses is genuine.
However, from our results we see that variations of the quark-gluon vertex strength with temperature and/or chemical potential allow to describe quite different scenarios for the longitudinal screening mass while an extension of the truncation is required for the transverse screening mass.

\subsection{Chiral condensate}
\label{sec:chiral_condensate}

\begin{figure*}[tb]
 \begin{center}
  \includegraphics[width=0.45\textwidth]{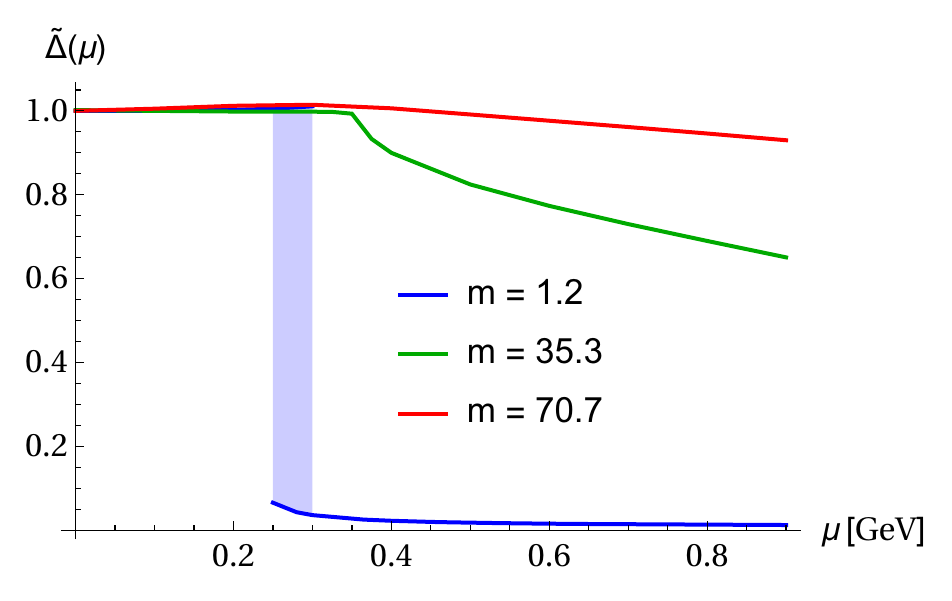}\hfill
  \includegraphics[width=0.45\textwidth]{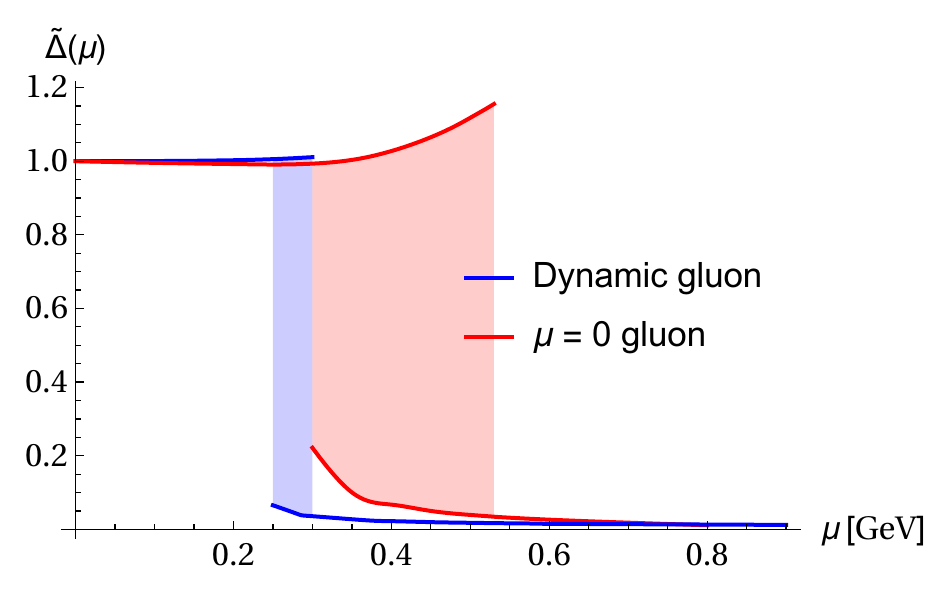}
  \caption{Left: Chiral condensate calculated with setup II of \tref{tab:parameters}.
  Right: Chiral condensate calculated with setup IIa of \tref{tab:parameters} using a fixed or a dynamic gluon propagator.}
 \label{fig:chiral_cond}
 \end{center}
\end{figure*}

The chiral condensate is a standard quantity to test for the breaking of chiral symmetry and thus to distinguish corresponding phases.
It can be calculated from the quark propagator as
\begin{align} \label{eq:chiralCondensate}
&\left<  \overline{\psi} \psi \right>  = -N_c Z_2 Z_m \sumint_q\Tr[S(q)]\nnnl
&\,=-N_c Z_2 Z_m \sum_{q_4}{\int{\frac{d^3q}{(2\pi)^3}\frac{4\,T\,B(q)}{A^2(q)\vec{q}^2 + C^2(q)q_4^2 + B^2(q)}}}.
\end{align}
Since this expression is UV divergent, it needs to be renormalized which is done by subtracting a quark condensate with a heavier renormalized mass $m_s$ from a condensate with a light renormalized mass $m_l$ which we take as the quark mass $m$ in this work:
\begin{align}
\Delta_{l,h}(\mu,T) = -\left<  \overline{\psi} \psi \right>_{l} + \frac{m_l}{m_s}\left<  \overline{\psi} \psi \right>_{h}.
\end{align}
It is also convenient to normalize this expression by the vacuum value
\begin{align}
 \widetilde{\Delta}(\mu,T)=\frac{\Delta_{l,h}(\mu,T)}{\Delta_{l,h}(0,0)}.
\end{align}

We show results for the chiral condensate and the different quark masses of setup II in the left plot of \fref{fig:chiral_cond}.
We observe the transitions at the same points and of the same types as for the longitudinal screening masses.
As we argued in Sec.~\ref{sec:gluon_prop}, it is quite likely that the gluon propagator does not depend strongly on the chemical potential.
Thus, we also tested the impact of the dynamic gluon propagator by comparing two calculations with a dynamic gluon propagator and with the gluon propagator fixed at $\mu=0$.
In the right plot of \fref{fig:chiral_cond}, one can see that the gluon propagator indeed influences the position of the transition.
For the future it will thus be important either to improve the setup for the gluon propagator or to confirm its weak dependence on the chemical potential in a certain range so that using the propagator from $\mu=0$ is a valid approximation.

\subsection{Quark number density}
\label{sec:quark_number_density}

The quark number density can be calculated from the quark propagator by
\begin{align}\label{eq:n}
n(\mu,T)=-\frac{\partial \Omega}{\partial \mu}=-N_c\, N_f\, Z_2\, \sumint_q \Tr[\gamma_4 \,S(q)],
\end{align}
where $\Omega$ is the grand-canonical potential of QCD given by
\begin{align}
 \Omega=-\frac{T}{V}\log{\mathcal{Z}(\mu,T)}.
\end{align}
$V$ is the volume of the system and $\mathcal{Z}$ the QCD partition function.
Eq.~\ref{eq:n} needs to be regularized which is done by subtracting a temperature independent term, see Refs.~\cite{Gao:2015kea,Gao:2016qkh,Isserstedt:2019pgx} for details:
\begin{align}\label{eq:n_reg}
 n_\text{reg}(\mu,T)=n(\mu,T) + N_c\, N_f\,Z_2 \int \frac{d^4q}{(2\pi)^4}\Tr[\gamma_4 \,S(q)].
\end{align}

\begin{figure}[tb]
 \begin{center}
  \includegraphics[width=0.45\textwidth]{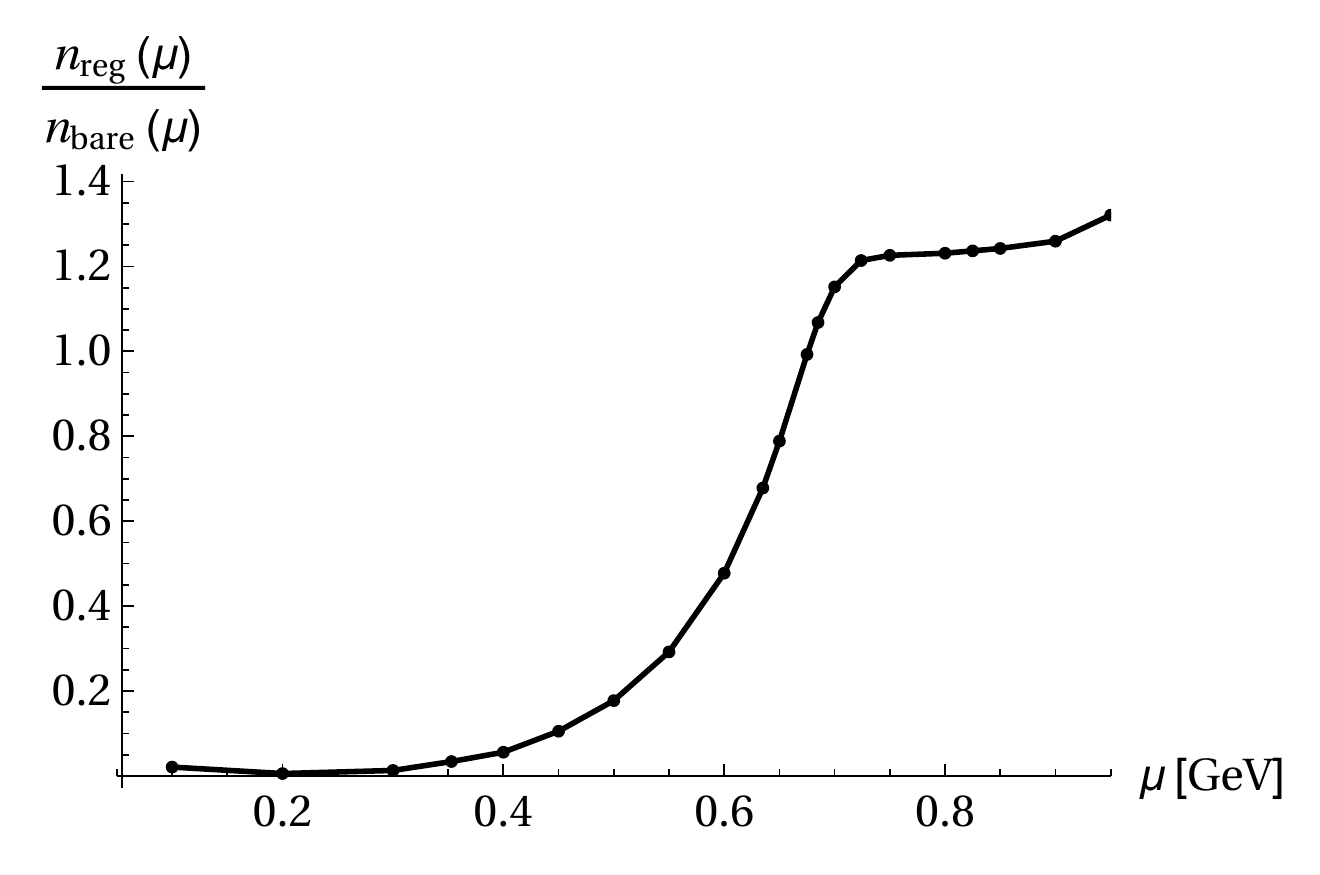}
  \caption{
  Quark number density calculated with setup I of \tref{tab:parameters}. $n_{bare}$ is the quark number density for a bare quark propagator.
  }
  \label{fig:qnd}
 \end{center}
\end{figure}

Results for the quark number density are shown in \fref{fig:qnd}.
Setup I was employed, viz., lattice data was used to fix the parameters.
Clearly, also in the quark number density a transition is seen at $\mu=700\,\text{MeV}$.

\subsection{Phase diagrams}
\label{sec:phase_diagrams}

Solving the present system of equations for higher temperatures, we can also investigate the crossover region and search for a critical point.
For comparison, we also show the same calculations for the gauge groups $SU(3)$ and $G_2$.
To connect to previous work, we use the setup of Ref.~\cite{Contant:2017gtz} which is detailed in \tref{tab:phase_diags}.
The setup for $SU(2)$ corresponds to setup IIa of \tref{tab:parameters}.
We recall that the value of the coupling is different for $G_2$ due to the choice of the gluon propagator; see Ref.~\cite{Contant:2017gtz} for details.
First results for $SU(2)$ and $G_2$ were already shown in Ref.~\cite{Contant:2017onc}.

\begin{table}[b]
 \begin{center}
  \begin{tabular}{|c||c|c|c||c|}
  \hline
    &$\alpha(\mu)$  & $d_1 [\text{GeV}^2]$ & $m\,[\text{MeV}]$ & $(\mu^\text{cep},T^\text{cep}) [\text{GeV}]$ \\
   \hline\hline
   $SU(3)$ & 0.3 & 7 & 1.2 & (0.165, 0.135)\\
   \hline
   $SU(2)$ & 0.3 & 15 & 1.18 & (0.172, 0.175)\\
   \hline
   $G_2$ & 0.45 & 6.78 & 1.2 & (0.156, 0.121)\\
   \hline
  \end{tabular}
 \caption{Couplings, quark-gluon vertex IR strength parameters, and quark masses used for the calculation of \fref{fig:phase_diags} and the resulting positions of the critical points.}
 \label{tab:phase_diags}
 \end{center}
\end{table}

For the confinement/deconfinement transitions the dual condensate \cite{Bilgici:2008qy,Synatschke:2007bz,Fischer:2009wc} is used, which is an order parameter for center symmetry in quenched QCD.
It is computed by introducing generalized $U(1)$ valued boundary conditions for the quarks $\psi(x, 1/T) = e^{i \varphi } \psi(x, 0)$ \cite{Fischer:2009wc} and projecting out the loops with winding number $1$:
\begin{align}
 \Sigma=\int_0^{2\pi}\frac{d\varphi}{2\pi}e^{-i\,\varphi}\left<  \overline{\psi} \psi \right>_\varphi d\varphi.
\end{align}
For the identification of the chiral transitions, the chiral condensate is used.
As definition for the transitions we use the  maxima of the following derivatives:
\begin{align}
 \chi_\text{ch} &= \frac{\partial \Delta_{l,h}}{\partial T},\\
 \chi_\text{dec} &= \frac{\partial \Sigma}{\partial T}.
\end{align}

\begin{figure}[tb]
 \begin{center}
 \includegraphics[width=0.45\textwidth]{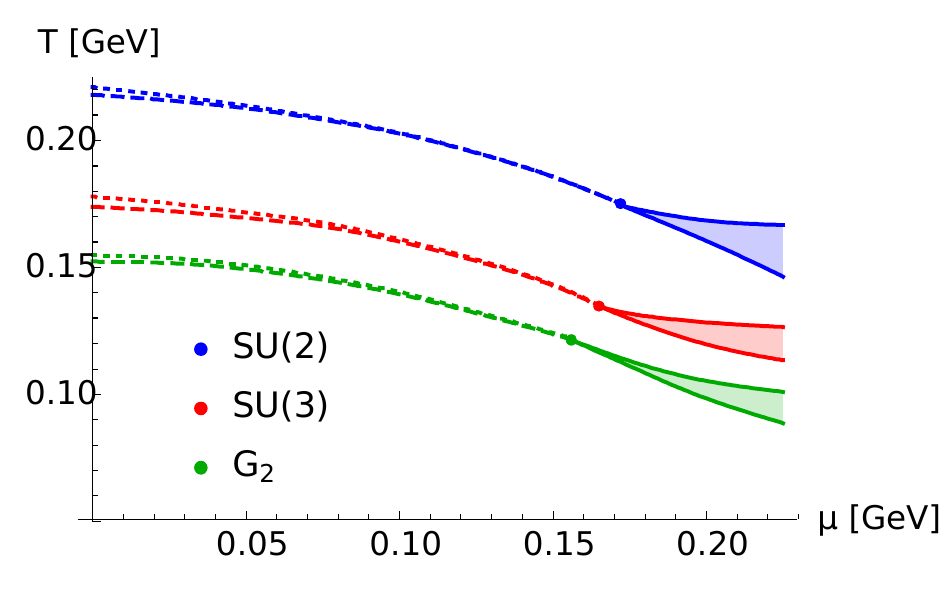}
 \caption{
 Phase diagrams for $SU(3)$, $SU(2)$, and $G_2$.
 The dashed lines represent the chiral transitions, the dotted lines the confinement/deconfinement transitions, and the continuous lines beyond the critical points are the spinodal lines of a first order transition.
 The employed setups are given in \tref{tab:phase_diags}.
 }
  \label{fig:phase_diags}
\end{center}
\end{figure}

The results for the chiral and confinement/deconfinement transitions are shown in \fref{fig:phase_diags}.
Clearly, all three theories show a similar behavior with a crossover from vanishing chemical potential to a critical point at $0.15\,\text{GeV}<\mu<0.175\,\text{GeV}$.
The lines of the chiral and confinement/deconfinement transitions are very close and merge at the critical points.
Their locations are given in \tref{tab:phase_diags}.
For all three theories, a first order transition is found beyond the critical points.

\section{Summary, conclusions and outlook}
\label{sec:summary}

We explored the phase diagram of QC$_2$D and compared various quantities with results from the lattice.
The employed truncation, which includes direct unquenching effects, uses a model for the quark-gluon interaction and fits for the (quenched) gluon propagator.
For low chemical potential, the results agree well with those from the lattice.
However, for higher chemical potential, we see deviations, which can be traced back to several sources.

One is related to the gluon propagator, for which we see a somewhat stronger dependence on chemical potential than expected from corresponding lattice results.
This, however, can easily be circumvented by using the unquenched gluon propagator from $\mu=0$.
This modification improves the agreement to some extent but is itself insufficient.
One possible source for these deviations is the quark-gluon vertex, which contains only a minimal dependence on temperature and density via the quark dressing functions in the model.
It was observed already earlier for $\mu=0$ that the model parameters need to be changed slightly when describing vacuum physics instead of the crossover at nonvanishing temperature.
Thus, it does not come as a surprise that a dependence on the chemical potential is missing as well.
To test this, we demonstrated successfully that a modified IR strength of the interaction leads to better agreement with lattice results.

\begin{figure}[tb]
 \begin{center}
 \includegraphics[width=0.45\textwidth]{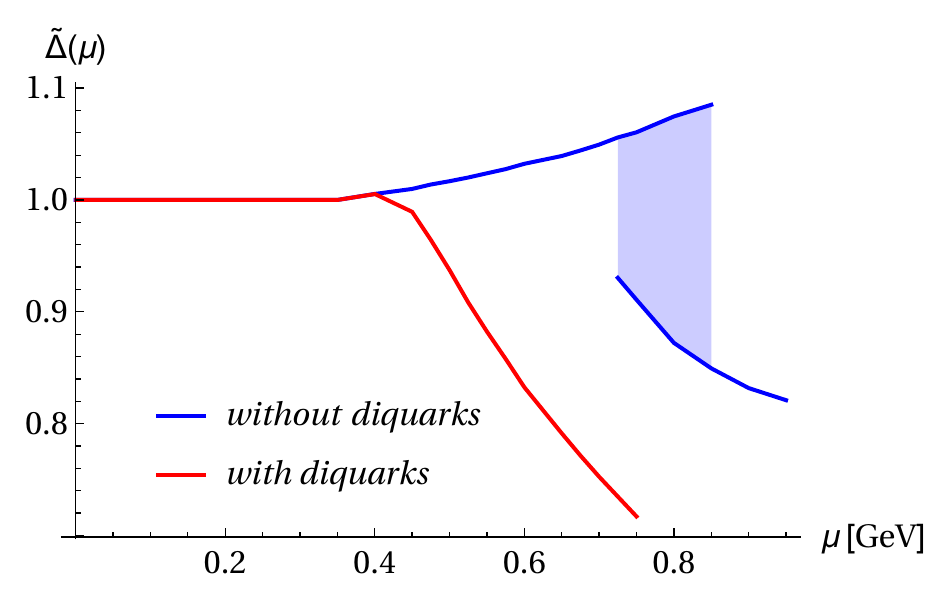}
 \caption{
 The chiral condensate with and without diquark contributions.
 The employed setup uses $d_1 = 10\, \text{GeV}^2$ and $m = 60\,\text{MeV}$.
 }
 \label{fig:chiral_cond_diquark}
\end{center}
\end{figure}

\begin{figure}[tb]
 \begin{center}
 \includegraphics[width=0.45\textwidth]{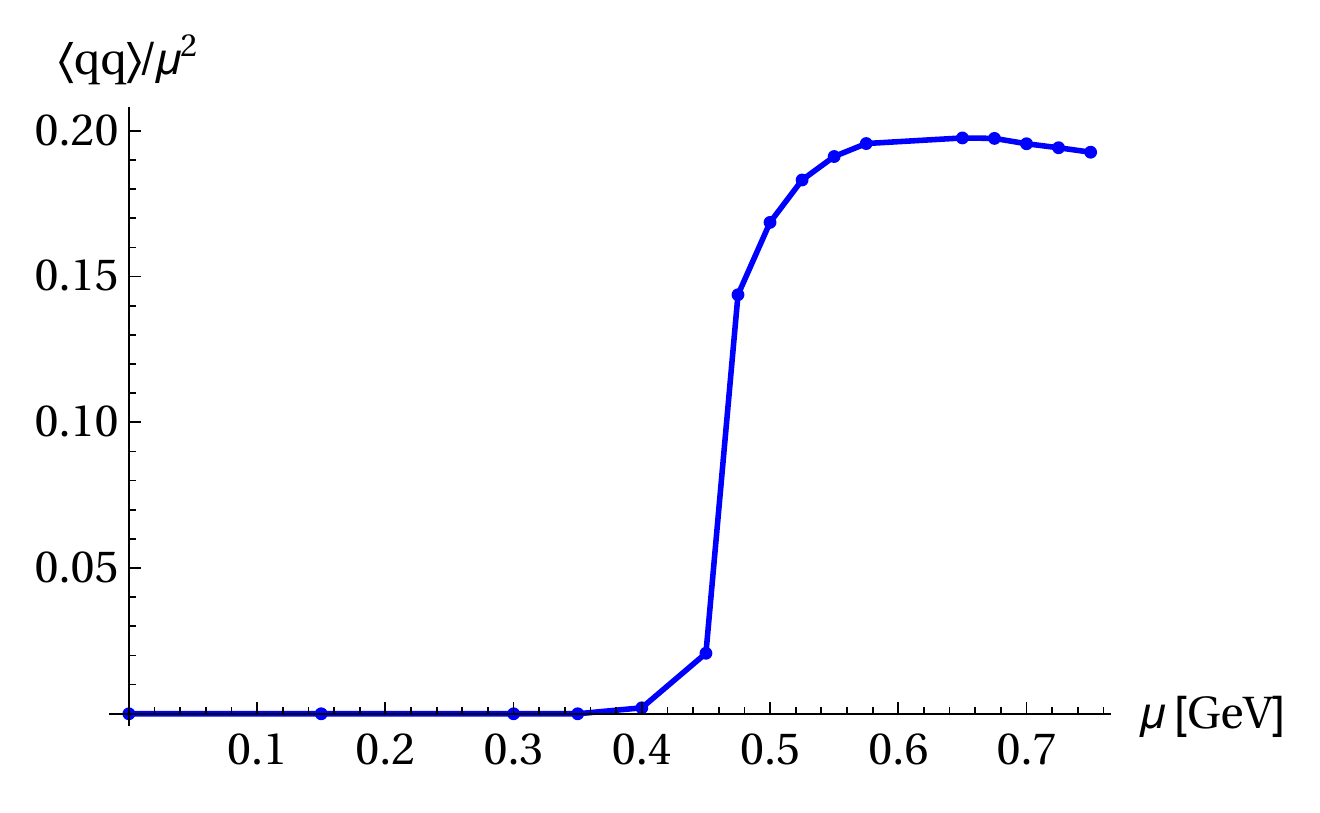}
 \caption{ 
 The diquark condensate as a function of chemical potential.
 }
\label{fig:diquark}
\end{center}
\end{figure}

Another reason for deviations from lattice results is that we did not include diquarks yet, which are expected to play an important role.
They can be included with the Nambu-Gor'kov formalism  \cite{Rischke:2003mt,Nickel:2006vf,Buscher:2014ixt,Muller:2016fdr} which leads to a more complex structure of the quark propagator,
\begin{align}
\mathcal{S}_\text{NG}(p) = \left(\begin{array}{cc} S^{+}(p) &  T^-(p) \\ T^+(p) & S^-(p) \end{array}\right).
\end{align}
$S^{+}(p)$ is the quark propagator from \eref{eq:quarkprop} and $S^{-}(p)$ is related to $S^{+}(p)$ by charge conjugation.
For details of the anomalous propagators $T^{\pm}(p)$, we refer to Ref.~\cite{Buscher:2014ixt}.
The diquark condensate, which is an order parameter for the diquark condensation phase, can be calculated as
\begin{align}
&\left<  q q \right>  = - Z_2 \sumint_q\Tr[\gamma_5 M T^{-}(q)],
\end{align}
where $M = T_2 \tau_2$ with $T_2$ ($\tau_2$) carrying the color (flavor) structure \cite{Kogut:2000ek}.

To get a first idea of the influence of diquarks we calculated for a few values of the chemical potential the chiral condensates with and without diquarks.
To compare with previous results, we choose the simplest generalization of the quark-gluon vertex from \eref{eq:qglvertModel} to the Nambu-Gor'kov setting by neglecting its anomalous components.
As gluon propagator we use the one fixed at $\mu=0$.
The result for the chiral condensate is shown in \fref{fig:chiral_cond_diquark}.
Not only does the position of the transition move, also the order changes: it is second order now.
The corresponding diquark condensate is shown in \fref{fig:diquark}.
As expected, it starts to increase at the same point at which the chiral condensate starts to drop.
For now, the calculation with diquarks has a limited resolution and 
more work is required, but they should be included in future work.

In summary, the setup of two-color QCD presents an interesting testbed where we can compare continuum with lattice methods.
Such comparisons are useful to show the path for future extensions of truncations of functional equations.
In particular since state-of-the-art truncations are quite demanding, such hints are valuable.

We also calculated the phase diagrams for QCD and QCD with the gauge group $G_2$ with the same truncation and found that the truncation behaves very similarly in all three cases.
This hints at a universal behavior which can be exploited by using QCD-like theories as a guide to improve truncations for real QCD.

One of the main results of this work is that the interaction between quarks and gluons as described by the quark-gluon vertex needs to be refined for large densities (and low temperatures) in order to achieve better precision.
However, the vertex is still quite an elusive object and information about its behavior beyond the vacuum is scarce.
First steps toward such a calculation were made only recently \cite{Welzbacher:2016rcv,Contant:2018zpi}. 
On the other hand, we can directly infer from the lattice results that the gluon propagator is most likely not what we need to be concerned with: its dependence on chemical potential is rather small.
This is reassuring, because a quantitative calculation of the propagator requires quite an elaborate truncation.
Thus, at the present level of precision, it is a convenient workaround to employ the gluon propagator from lattice calculations.

\section{Acknowledgments}
We thank Axel Maas and Ouraman Hajizadeh for useful discussions and for providing lattice data.
We are grateful to Christian Fischer and Philipp Isserstedt for helpful discussions.
HPC Clusters at the University of Graz were used for the numerical computations.
The software programs and packages \textit{Mathematica} \cite{Wolfram:2004}, \textit{DoFun} \cite{Alkofer:2008nt,Huber:2011qr,Huber:2019dkb}, and \textit{CrasyDSE} \cite{Huber:2011xc} were used for deriving and solving numerically the DSEs.
Feynman diagrams were created with Jaxodraw \cite{Binosi:2003yf}. 
Support by the FWF (Austrian science fund) under Contract No. P27380-N27 and through the doctoral program ``Hadrons in Vacuum, Nuclei and Stars'', Contract W1203-N16, is gratefully acknowledged.

\bibliographystyle{utphys_mod}
\bibliography{literature_quSU2AtTmu}

\end{document}